\documentstyle[12pt]{article}

\def\beq{\begin{equation}}
\def\eeq{\end{equation}}
\def\bea{\begin{eqnarray}}

\def\eea{\end{eqnarray}}
\def\ds{\displaystyle}

\def\ssz{\scriptsize}

\def\ni{\noindent}

\def\req#1{(\ref{#1})}

\def\rep{\mbox{Re}\ }

\def\aR{\mbox{\bf R}}
\def\Zed{\mbox{\bf Z}}

\def\barc{\begin{array}{c}}
\def\ear{\end{array}}
\def\lsect#1{\vspace{5mm} {\ni\bf #1.}}
\def\cam{}

\newcommand{\ee}{\end{equation}}
\newcommand{\be}{\begin{equation}}

\newcommand{\ea}{\end{eqnarray}}
\newcommand{\ba}{\begin{eqnarray}}

\begin{document}


\ni{\Large\bf
Energy of the vacuum with a perfectly conducting 
and infinite cylindrical surface
}

\vspace*{5mm}
\ni Peter Gosdzinsky$^{a}$
and August Romeo$^{b}$
\\
\ni ${}^a$ {\it Universitat Aut{\`o}noma de Barcelona and IFAE,
Edifici Cn, E-08193 Bellaterra} \\
\ni ${}^b$ {\it IEEC (Institut d'Estudis Espacials de Catalunya),
CSIC Research Unit,
Edifici Nexus, 204, c. Gran Capit{\`a} 2-4, E-08034 Barcelona}
\vspace*{5mm}

\ni{\bf Abstract.}
Values for the vacuum energy of scalar fields under Dirichlet and Neumann
boundary conditions on an infinite clylindrical surface are found,
and they happen to be of opposite signs.
In contrast with classical works, a complete zeta function
regularization scheme is here applied.
These fields are regarded as interesting both by themselves and
as the key to describing the electromagnetic (e.m.) case.
With their help, the figure for the e.m. Casimir effect
in the presence of
this surface,
found by De Raad and Milton, is now confirmed.

\ni{\small PACS: 03.70.+k, 12.20.-m, 42.50.Lc}

\vspace*{5mm}

\lsect{Introduction}
The Casimir effect \cite{Cas} is caused by
zero-point fluctuations of quantum fields in the vacuum when
they are modified by the introduction of boundaries or constraints.
The particular features of the resulting force depend on
the nature of the field, boundaries and boundary conditions (b.c.) imposed
(in some cases, an interpretation in terms of the radiation
pressure \cite{MCC} has been made).
It
arises in different areas of physics, including quantum electrodynamics,
theory of hadrons, and cosmology. 
Further, there has
been a series of attempts to explain the phenomenon of sonoluminiscence
on the basis of the Casimir effect, which, although unsuccessful 
(see \cite{MNson} and refs. therein) still gives an idea of the importance
attributed to its role in modern physics.

Evaluating the zero-point energy is usually an involved problem, and a
great deal of techniques have been thought up for this purpose.
Some reasons have made of zeta function
regularization\cite{BM}-\cite{EORBZ} a relatively popular method.
Particularly, the variant we employ in this paper has similarities with
the technique developed in ref.\cite{BKK} for a cosmological problem.
In ref. \cite{LRap}, it was successfully applied to the cases of the sphere
and the circle reobtaining known results \cite{Bo}-\cite{MN} and finding
some new ones.
Old as the spherical problem may look, it has been repeatedly
revisited in such recent papers as refs. \cite{NP,BH}.
However, as far as we know, a complete zeta function approach
has never been applied to the problem of the cylinder,
first studied in ref.\cite{MiRa}.
This is precisely what we plan to do in the present work.

The spectral zeta functions, made from some operator's eigenmode set
$\{ \omega_n \}\equiv {\cal M} $, will be called
\beq
\zeta_{{\cal M}}(s)=\sum_n \omega_n^{-s}.
\label{defZMod}
\eeq
Quite often, it is even more convenient to use the dimensionless version
$\ds
\zeta_{{\cal M}\over \mu}(s)=
\sum_n \left( \omega_n\over \mu \right)^{-s},
$
where $\mu$ is an arbitrary scale with mass dimensions.
In most cases of interest, we have to consider spectra of the same sort as
that of a free particle, and $\omega_n$ grows with $n$ without bound.
Therefore, strictly speaking, these identities hold only for
$\rep s > s_0$, being $s_0$ a positive number given by the rightmost
pole of $\zeta_{{\cal M}}(s)$. Nevertheless, this function admits
analytic continuation to other values of $s$,
including the negative reals.

The vacuum energy density $\varepsilon$
will be obtained by zeta-regularization of the mode sum
$\ds {1\over 2}\sum_n \omega_n$
(we have adopted the typical QFT units, i.e., $\hbar=c=1$).
If the result found is not finite yet, one can add
a principal part (P) prescription as in ref.\cite{BVW}. Writing
everything together,
\begin{equation}
\varepsilon(\mu)= \mbox{P}\left[
{1 \over 2} \mu \, \zeta_{{\cal M}\over \mu}(s); s\to -1
\right] ,
\label{PPP}
\end{equation}
where P means extraction of the principal part.
Obviously, for this procedure to be operational, we need
the analytic continuation of $\zeta_{{\cal M}}(s)$
to $s=-1$.
In our case, as we have an infinite cylinder, we
consider $\varepsilon$ as the `linear' density of energy per length unit.

The present problem can be related to that of a circle in $2+1$
dimensions. If $\{ \omega_n \}\equiv {\cal C}$ denotes the set of
all the eigenfrequencies, the complete zeta function for the circle will be
$\ds\zeta_{\cal C}(z)=\sum_{\omega_n \in {\cal C}} \omega_n^{-z}$.
Given that the ones for the cylinder are
${\cal M}=\{ \sqrt{ \omega_n^2+k^2 } , \omega_n\in{\cal C}, k\in\aR \}$, the
new zeta function will be
\cam
\beq
\zeta_{{\cal M}}(s)=\int_{-\infty}^{\infty} {dk\over 2\pi} \,
\sum_{\omega_n \in {\cal C}}
\left( \sqrt{ \omega_n^2+k^2 } \right)^{-s} = {1 \over {2 \pi}}
B\left( {1 \over 2}, {s-1\over 2} \right) \zeta_{\cal C}(s-1),
\eeq
where $B$ stands for the Euler beta function.
Laurent-expanding around $s=-1$ one realizes that
\cam
\beq
%
\zeta_{{\cal M}}(s)={1 \over 2\pi}\left[
{ \zeta_{\cal C}(-2) \over s+1 }
+\bigg( \ln(2)- {1 \over 2} \bigg) \zeta_{\cal C}(-2)
+\zeta_{\cal C}'(-2)
+{\cal O}(s+1) \right] .
\eeq
Thus, to determine $\zeta_{{\cal M}}(s)$ at $s=-1$ means having to know
$\zeta_{\cal C}(z)$ at and around $z=-2$. For this reason, it is not
possible to directly apply the results in ref.\cite{LRap} for the circle,
which only give information about $\zeta_{\cal C}(z)$ at $z=-1$.
The residue of $\zeta_{{\cal M}}(s)$ at $s=-1$ is
$\zeta_{\cal C}(-2)$ itself. In fact, it turns out that
the sum of the internal and external contributions to this quantity vanishes.
This can be anticipated by the parity of the Seeley - de Witt coefficient
related to the possible divergence (actually, the same as the coefficient 
for the scale anomaly of the one-loop energy)
and by the fact that we are including the
internal and external spaces
to an infinitely thin boundary in a globally flat manifold of odd space 
dimension ---see e.g. \cite{BVW}, \cite{EORBZ} (pages 107-111),
\cite{BS} or refs. therein.
Such a vanishing means two things: 
\begin{itemize}
\item First, the whole result is finite
and the principal part prescription, 
as well as the use of the scale $\mu$, are actually unnecessary.
Then, \req{PPP} reduces to
\beq
\varepsilon={1 \over 2}\zeta_{\cal M}(-1).
\label{noPP}
\eeq
\item Second, the finite value which appears at the end depends only on
$\zeta_{\cal C}'(-2)$, i.e. \cam
\beq
\zeta_{\cal M}(s)={1 \over 2\pi}\zeta_{\cal C}'(-2) + {\cal O}(s+1).
\label{zcylzcirc1}
\eeq
Therefore, to know only the value
of $\zeta_{\cal C}$ at $s=-2$ ---without knowing the next
term in its Taylor expansion--- would be of no help.
\end{itemize}

The present paper is organized as follows: In the next section we
consider the Maxwell equations and how the e.m. problem decomposes
into two situations with just scalar fields. Afterwards, the relevant
zeta functions are introduced in sec. 3. Then, the energetic contributions
of the Dirichlet and Neumann modes are separately found in secs, 4 and 5, 
respectively. The total result and related comments appear in sec. 6.

\lsect{Solutions of the Maxwell equations for a cylinder}
\cam
The eigenfrequency set of an electromagnetic problem is the
one dictated by the Maxwell equations under the b.c. in question.
Therefore, our first step will be to solve the Maxwell equations.
We will look for solutions in cylindrical coordinates of the form
\be
E_i, B_j = e^{i \omega t} e ^{i k z} e ^{im \theta} R(r).
\label{modes}
\ee

We observe that in \req{modes}
$\omega > k$, which allows us to boost to a frame in which $k=0$. 
In this frame, Maxwells equations read
\be
\left. 
\begin{array}{l}
{\displaystyle {im \over r} B_z = -i \omega E_r} \\
{\displaystyle -{\partial B_z \over \partial r} = i \omega E_\theta } 
\\
{\displaystyle  {\partial (r B_\theta)  \over \partial r} -i m B_r
= i \omega r E_z }
\end{array} \right\}
\qquad{\mbox{and}} \qquad
\left. 
\begin{array}{l}
{\displaystyle {im \over r} E_z = i \omega B_r} \\
{\displaystyle -{\partial E_z \over \partial r}= -i \omega B_\theta } 
\\ 
{\displaystyle  {\partial (r E_\theta)  \over \partial r} -i m E_r
= -i \omega r B_z }
\end{array} \right\}
\label{maxdec}
\ee
In fact, \req{maxdec} are two decoupled systems of three equations
each, which can be solved using well known techniques. Inside the
cylinder, after imposing regularity at $r=0$
(which excludes the $Y_m$ solutions), we find \cam two sets of
solutions: The first one is given by
\be
B_z = J_m(\omega r), \qquad 
E_r = {m \over \omega r} J_m(\omega r), \qquad 
E_\theta = {i \over \omega} {\partial J_m(\omega r) \over \partial r}
\label{solu1}
\ee
while the second one reads
\be
E_z = J_m(\omega r), \qquad 
B_r = -{m \over \omega r} J_m(\omega r), \qquad 
B_\theta = -{i \over \omega} {\partial J_m(\omega r) \over \partial r}
\label{solu2}
\ee
where we only list the radial parts of the
solutions, and the components not listed vanish. 
Calling $\vec{n}$ the normal vector to the boundary,
the b.c. for an infinite and
perfectly conducting cylinder of radius $a$, which read
$\left. \vec{n}\cdot \vec{B} \right\vert_{r=a}=0$,
$\left. \vec{n}\times \vec{E} \right\vert_{r=a}=0$,
imply $E_{\theta}=E_z=0$, $B_r=0$, and, thus,
$J'_m (\omega a) = 0$
for \req{solu1}, and
$J_m (\omega a) = 0$
for \req{solu2}.

In regard to the solutions outside of the cylinder, we have to bear in mind
that the $Y_m$ functions are no longer to be ruled out. Thus,
we may start by replacing
$J_m(\omega r) \rightarrow A J_m(\omega r) + B Y_m(\omega r)$
in \req{solu1} and \req{solu2}. Next, the conditions at infinity
impose constraints on $A$ and $B$. This question has been
addressed in some detail in \cite{LRap} for the sphere, but the same type
of calculation may be carried over to the present case.
The outcome is that, in practice,
everything works out as if the actual combination was
$J_m(\omega r) +i Y_m(\omega r)=H^{(1)}(\omega r)$ (or its conjugate).
Once these solutions have been obtained, one just has to boost back
to the original frame.

Summing up,
the Maxwell equations for cylindrical waves under perfectly conducting
b.c. on an infinite cylinder of radius $a$
have two sets of solutions, corresponding to the
transverse electric (TE) and transverse magnetic (TM) modes.
TE modes are subject to Dirichlet b.c.,
and TM modes obey Neumann b.c.
Thus, when restricting the problem to the circle, the set of modes may be written as
${\cal C}={\cal D}\cup{\cal N}$, where ${\cal D}$ stands for the Dirichlet
modes and ${\cal N}$ for the Neumann modes.
These two sets are further subdivided, depending on whether we consider
the part of the solution inside or outside of the cylinder. In the first case,
we add the requirement of regularity in the interior, while
in the second we have to impose physically suitable conditions at infinity.
For the internal TE modes \req{solu2}, the $\omega$'s are such that $\omega a$
must be nonvanishing zeros
of the Bessel function $J_{m}$, i.e.,
\beq
J_{m}(\omega a)=0, \ m \in \Zed,
\label{Dicond}
\eeq
where $m$ is the angular momentum number,
while
the internal TM eigenfrequencies \req{solu1} obey a Neumann condition of the form
\beq
J_{m}'(\omega a)=0,
\ m \in \Zed.
\label{Necond}
\eeq
As for the external modes, and in view of the
the analytical methods we plan to apply, it is formally
enough to replace every Bessel function $J_m$ by a Hankel function
$H_m^{(1)}$ (justified by \cite{LRap}). In consequence,
$I_{m}$ functions should be replaced by $K_{m}$ functions 
\cam (See below).

\lsect{Partial-wave and complete zeta functions}
As we have just seen, the e.m. problem can be regarded as the sum of two situations
involving a massless scalar field: in the first, it is subject to \req{Dicond},
and, in the second, has to satisfy \req{Necond}.
Let's now consider an adequate formalism for each part.

First, we introduce the `partial-wave' zeta functions for the modes
inside the cylinder (which we shall call `int' region)
\beq
\zeta_{m}^{{\mbox{\ssz int}}, {\cal D}}(s)=
\sum_{n=1}^{\infty} j_{m, n}^{-s} \, ,
\hspace{0.5cm}
\zeta_{m}^{{\mbox{\ssz int}}, {\cal N}}(s)=
\ds\sum_{n=1}^{\infty} k_{m, n}^{-s} \, ,
\hspace{1cm}
\mbox{for $\rep s > 1$}, \label{zetanuSe}
\eeq
where $j_{m, n}$ is the $n$th nonvanishing zero of $J_{m}$ and
$k_{m, n}$ means the $n$th solution of eq. \req{Necond}, for a given
$m$.
The superscripts ${\cal D}$ and ${\cal N}$ refer to Dirichlet and Neumann b.c.
The analogous functions for the external modes (`ext' region) will be called
$
\zeta_{m}^{{\mbox{\ssz ext}}, {\cal D}}(s) , 
\zeta_{m}^{{\mbox{\ssz ext}}, {\cal N}}(s) . 
$
Looking at the whole problem,
we have to take into account the angular degeneracies
\beq
d(m)=\left\{
\begin{array}{ll}
1,&\mbox{for $m = 0$,} \\
2,&\mbox{for $m > 0$.}
\end{array}
\right.
\eeq
With them, we define the complete zeta functions for the circle:
\begin{equation}
\begin{array}{lllll}
\ds\zeta_{\cal C}^{{\mbox{\ssz int}}, {\cal D}}(s)&=&
\ds a^s \sum_{m=0}^{\infty} d(m)
\sum_{n=1}^{\infty} j_{ m, n}^{-s}
&=&\ds a^s \sum_{m=0}^{\infty} d(m) \,
\zeta_{m}^{{\mbox{\ssz int}}, {\cal D}}(s) , \\
\ds\zeta_{\cal C}^{{\mbox{\ssz int}}, {\cal N}}(s)&=&
\ds a^s \sum_{m=0}^{\infty} d(m)
\sum_{n=1}^{\infty} k_{ m, n}^{-s}&=&
\ds a^s \sum_{m=0}^{\infty} d(m) \,
\zeta_{m}^{{\mbox{\ssz int}}, {\cal N}}(s) ,
\end{array}
\label{defzdDs}
\end{equation}
and
\beq
\begin{array}{lll}
\ds\zeta_{\cal C}^{{\mbox{\ssz ext}}, {\cal D}}(s)
&=&\ds a^s \sum_{m=0}^{\infty} d(m) \,
\zeta_{m}^{{\mbox{\ssz ext}}, {\cal D}}(s) , \\
\ds\zeta_{\cal C}^{{\mbox{\ssz ext}}, {\cal N}}(s)
&=&\ds a^s \sum_{m=0}^{\infty} d(m) \,
\zeta_{m}^{{\mbox{\ssz ext}}, {\cal N}}(s) .
\end{array}
\label{defzdDsII}
\eeq
With these ingredients, one builds up the total complete zeta function
for the circle, i.e. the one entering eq.\req{zcylzcirc1}, which is
\beq
\begin{array}{cccccc}
\zeta_{\cal C}(s)&=&
\zeta_{\cal C}^{{\mbox{\ssz int}}, {\cal D}}(s)
+\zeta_{\cal C}^{{\mbox{\ssz ext}}, {\cal D}}(s)
&+&\zeta_{\cal C}^{{\mbox{\ssz int}}, {\cal N}}(s)
+\zeta_{\cal C}^{{\mbox{\ssz ext}}, {\cal N}}(s) \\
&\equiv& \zeta_{\cal C}^{\cal D}(s) &+& \zeta_{\cal C}^{\cal N}(s) .
\end{array}
\eeq

By the methods of ref.\cite{LRap}
the functions \req{zetanuSe} have
the following integral representations
valid for $-1 < {\rm Re}\  s < 0$:
\beq
\begin{array}{c}
\ds\zeta_{m}^{{\mbox{\ssz part, cond}}}(s)={s \over \pi} \sin{\pi s \over 2}
\int_0^{\infty} dx \, x^{-s-1} \,
\ln\left[ L^{{\mbox{\ssz part, cond}}}(m, x) \right] , \\
\mbox{ part $\in \{ $ int, ext $ \} $, cond $\in \{ {\cal D}, {\cal N} \}$, }
\label{reprzetanuIR}
\end{array}
\eeq
where
\beq
\begin{array}{llllll}
L^{{\mbox{\ssz int}}, {\cal D}}(m ,x)&=&\ds\sqrt{2\pi x} \, e^{-x} \, I_{m}(x),  \qquad
L^{{\mbox{\ssz int}}, {\cal N}}(m ,x)&=&\ds\sqrt{2\pi x} \, e^{-x} \, I_{m}'(x), \\
L^{{\mbox{\ssz ext}}, {\cal D}}(m ,x)&=&\ds\sqrt{2x \over\pi } \, e^{x} \, K_{m}(x), \qquad
L^{{\mbox{\ssz ext}}, {\cal N}}(m ,x)&=&\ds-\sqrt{2x \over\pi} \, e^{x} \, K_{m}'(x) .
\end{array}
\eeq
This is not enough yet, as we must go further to the left, until $s=-2$.
In order to extend the domain of validity of these formulas, it suffices
to perform an adequate subtraction in the integrand, with the aim of
separating the part that causes the divergent behaviour which stops us from
reaching $s=-2$.
One
\cam of the simplest possibilities is to add and subtract functions of
the type
\beq
x^{-s-1} [ {\cal Y}_1(m) t(x)+ {\cal Y}_2(m) t^2(x) ],
\qquad
t(x)=(1 + x^2)^{-1/2}
\label{deftx}
\eeq
Specifically, we take
\cam
\beq
\begin{array}{lllllll}
{\cal Y}_1^{\mbox{\ssz int},{\cal D}}(m)&=&
\ds -{1 \over 2}\left( m^2 -{1\over 4} \right),
&\hspace{0.5cm}&
\ds {\cal Y}_2^{\mbox{\ssz int},{\cal D}}(m)&=&
\ds -{1 \over 4}\left( m^2 -{1\over 4} \right), \\
{\cal Y}_1^{\mbox{\ssz int},{\cal N}}(m)&=&
\ds -{1 \over 2}\left( m^2 +{3\over 4} \right),
&\hspace{0.5cm}&
{\cal Y}_2^{\mbox{\ssz int},{\cal N}}(m)&=&
\ds +{1 \over 4}\left( m^2 -{3\over 4} \right), \\
{\cal Y}_1^{\mbox{\ssz ext, cond}}(m)&=& 
-{\cal Y}_1^{\mbox{\ssz int, cond}}(m),
&\hspace{0.5cm}&
{\cal Y}_2^{\mbox{\ssz ext, cond}}(m)&=& 
{\cal Y}_2^{\mbox{\ssz int, cond}}(m),
\hspace{0.5cm} \mbox{cond $\in \{ {\cal D}, {\cal N} \}$,}
\end{array}
\eeq
\cam
and write
\beq
\begin{array}{c}
\ds\zeta_{m}^{{\mbox{\ssz part, cond}}}(s)={s \over \pi} \sin{\pi s \over 2} \\
\ds\times \left\{
\int_0^{\infty} dx \, x^{-s-1} \bigg[
\ln\left[ L^{{\mbox{\ssz part, cond}}}(m, x) \right]
-{\cal Y}_1^{{\mbox{\ssz part, cond}}}(m)t(x)
-{\cal Y}_2^{{\mbox{\ssz part, cond}}}(m)t^2(x) \bigg] \right.  \\
\ds\left.
+{\cal Y}_1^{{\mbox{\ssz part, cond}}}(m){1 \over 2}
B\left( {s+1 \over 2}, -{s \over 2} \right)
+{\cal Y}_2^{{\mbox{\ssz part, cond}}}(m){1 \over 2}
B\left( {s+2 \over 2}, -{s \over 2} \right)
\right\} , \\
\mbox{ part $\in \{ $ int, ext $ \} $, cond $\in \{ {\cal D}, {\cal N} \}$. }
\label{reprzetanu0}
\end{array}
\eeq
Given that all the represented integrands have now the asymptotic behaviour
$x^{-s-1}\cdot{\cal O}\left( t^3(x) \right)$,
the integrals are finite at $s=-2$. 
At this point, their specific values do not matter,
as they are eventually multiplied by the zero coming from $\sin{\pi s\over 2}$.
Further, since
$\ds\lim_{s\to -2}\sin{\pi s \over 2}
B\left( {s+2 \over 2}, -{s\over 2} \right)=-\pi$, all the expressions have
a finite limit at $s=-2$. For instance,
\beq
\ds\zeta_{m}^{{\mbox{\ssz int}}, {\cal D}}(-2)=
\ds -{1 \over 4}\left( m^2 - {1 \over 4} \right), \qquad
\ds\zeta_{m}^{{\mbox{\ssz int}}, {\cal N}}(-2)=
\ds {1 \over 4}\left( m^2 - {3 \over 4} \right) .
\eeq
The first confirms a previous result in ref.\cite{LRjpa}, while the
other is a new result.

\lsect{Dirichlet modes}
\cam
The integral in eq. \req{reprzetanu0} converges for $s=-2$ and can be
obtained numerically. In order to compute the complete $\zeta$
functions, \req{defzdDs} and \req{defzdDsII},
it proves convenient to perform further manipulations. This job
has already been done in ref.\cite{LRap}. Although the relevant $\zeta$
functions of that paper were obtained starting from expresions that 
are valid for $-1 < {\rm Re}~ s < 0$, we can make use of their final
expresion (eq.(3.22) of ref.\cite{LRap} with $\sigma _1^{I, \cal D}
=-\sigma _2^{I, \cal D}=-1$ for the interior modes,  and
$\sigma _1^{II, \cal D}= \sigma _2^{II, \cal D} = -1$ for the exterior
modes)  provided  $N>2$. Adding the contribution of the internal and
external modes, we obtain
\ba
\zeta_{m}^{\cal D}(s)&=&
m^{-s} \bigg\{ -{1 \over 2} - {s^2(6+5s) \over 128 m ^2}
+{s^2(2+s)(176-268 s -452 s^2 - 113 s^3)\over 98304 m ^4} \bigg\}
\nonumber \\
&&+{s \over \pi}\sin {\pi s \over 2} m ^{-s} {\cal S}_4^{\cal D}(s, m) .
\label{theZETAdir}
\ea
where we have defined
\beq
\begin{array}{c}
\ds\int_0^{\infty} dx \, x^{-s-1} \, \left\{
\ln\left[ 2m \sqrt{1+x^2} I_{m}(m x) K_{m}(m x) \right]
-2\sum_{2 \le 2n \le N} { {\cal U}^{{\cal D}}_{2n}(t(x)) \over m^{2n} }
\right\}
\equiv {\cal S}^{{\cal D}}_N(s, m) .
\label{SDNsnu}
\end{array}
\eeq
and \cam
\begin{equation}
\begin{array}{lll}
\ds{\cal U}^{{\cal D}}_2(t)&=&\ds
{{{t^2}}\over {16}} - {{3\,{t^4}}\over 8} +
 {{5\,{t^6}}\over {16}}, \\
\ds{\cal U}^{{\cal D}}_4(t)&=&\ds
     {{13\,{t^4}}\over {128}} -
     {{71\,{t^6}}\over {32}} +
     {{531\,{t^8}}\over {64}} -
     {{339\,{t^{10}}}\over {32}} +
     {{565\,{t^{12}}}\over {128}} , \\
&\vdots&
\end{array}
\label{U14}
\end{equation}
\cam
In \req{theZETAdir} we have chosen $N=4$. For the coefficients
${\cal S}^{{\cal D}}_4(s, m)$ we find
{\small
\ba
{\cal S}_4^{\cal D}(-2,1) &=& 1.34757127 \times 10^{-3}, \quad
{\cal S}_4^{\cal D}(-2,2) = 2.83382470 \times 10^{-5}, \quad
{\cal S}_4^{\cal D}(-2,3) = 2.6944637 \times 10^{-6}, \nonumber\\
{\cal S}_4^{\cal D}(-2,4) &=& 4.9500306 \times 10^{-7}, \quad
{\cal S}_4^{\cal D}(-2,5) = 1.31795554 \times 10^{-7}, \quad
{\cal S}_4^{\cal D}(-2,6) = 4.4524310 \times 10^{-8}, \nonumber \\
{\cal S}_4^{\cal D}(-2,7) &=& 1.7751724 \times 10^{-8}, \quad
{\cal S}_4^{\cal D}(-2,8) = 7.99494 \times 10^{-9}, \qquad \cdots
\label{numBIGSN}
\ea
}
When $m$ is large enough,  ${\cal S}_4^{\cal D}(-2,m)$ decreases as
$m^{-6}$, and can be approximated by
\be
{\cal S}_4^{\cal D}(-2,m) = {19 \over 8960 m^6} -
{ 7649 \over 4730880 m^8} +{192349 \over 82001920 m^{10}}
-{15293983 \over 2788065280 m^{12}} .
\ee
In order to construct the complete zeta function, we need the series
\ba
\sum_{m=1}^\infty \zeta_{m}^{\cal D}(s)&=&-{1 \over 2} \zeta_R(s)
-{s^2(6+5s) \over 128} \zeta_R(2+s)
+{s^2(2+s)(176-268 s -452 s^2 -113 s^3) \over 98304} \zeta_R(4+s)
\nonumber \\
&&+{s \over \pi} \sin { \pi s \over 2}
\sum_{m=1}^\infty m^{-s} {\cal S}_4^{\cal D}(s,m) ,
\label{sumZETA}
\ea
where the appearance of the Riemann zeta functions $\zeta_R$ comes
from identifying sums of the type $\ds\sum_{m=1}^{\infty}m^{-z}=\zeta_R(z)$.
Observe that, strictly speaking, such identification is only valid
if $\rep(z) > 1$. Given that we are analytically continuing expressions
involving $s$, we have to suppose that we may temporarily take $s$ large enough
so that this process be valid, before letting $s \to -2$ at the end.
It appears also the series
\be
\sum_{m=1}^\infty m^2 {\cal S}_4^{\cal D}(-2,m) = 
0.001500509798
\ee
Expression \req{theZETAdir} is not valid for $m=0$, since it was
obtained from a rescaling $x \to m x$ and
application of uniform asymptotic expansions in $m x$. However, the
$m=0$ contribution
can be easily found by applying formulas
\req{reprzetanu0} and preceding,
which yield
\be
\zeta_0^{\cal D}(s)=
\zeta_0^{{\mbox{\ssz int}}, {\cal D}}(s)+\zeta_0^{{\mbox{\ssz ext}}, {\cal D}}(s)=
-{s \over 16} + {s \over \pi} \sin { \pi s \over 2}
\int_0^\infty dx \, x^{-1-s} \, \bigg\{ \ln \{2 x I_0(x) K_0(x) \}
-{1 \over 8} {1 \over 1+x^2} \bigg\} .
\ee
Numerically, we find
\be
\int_0^\infty dx \, x \bigg\{ \ln \{2 x I_0(x) K_0(x) \}
-{1 \over 8} {1 \over 1+x^2} \bigg\}=
0.01096298110873
\label{intI0K0}
\ee
Putting everything in \req{sumZETA}-\req{intI0K0} together, 
we obtain, for $s \sim -2$,
\be
a^s \, \zeta_{\cal C}^{\cal D}(s)=
\sum_{m= 0}^\infty d(m) \zeta_m^{\cal D}(s) = 0.007725967 (s+2)
+{\cal O}\left( (s+2)^2 \right)
,\qquad
\ee
and, therefore,
\beq
\varepsilon^{\cal D} = {0.000614794033\over a^2} ,
\label{scalarENE}
\eeq
where $\varepsilon^{\cal D}$ means the part of $\epsilon$ coming from
the contribution of $\zeta_{\cal M}^{\cal D}$ only, and can be
interpreted as the Casimir energy linear density of a scalar field under
Dirichlet b.c. on the cylindrical surface. As expected, \req{scalarENE}
is free of divergences, which provides us with a nontrivial check of
our result. \cam The result in \req{scalarENE}  (and \req{laN} and
\req{totalDEN} below as well) is the lowest order result, which we
expect to be modified by higher order (quantum, finite temperature...)
effects.

\lsect{Neumann modes}
\cam
The Neumann modes admit a very similar treatment. For the equivalent of
\req{theZETAdir} we find
\be
\zeta ^{\cal N}_m (s)
= m^{-s} \left\{ {1 \over 2 } + {s^2 (2+7s)\over 128 m ^2}
+{s^2(s+2)(-16 +548 s +2156 s^2 +707 s^3) \over 491520 m ^4}
+{s \over \pi} \sin \bigg( {\pi s \over 2} \bigg)  
{\cal S}^{\cal N}_4(s,m) \right\}
\label{theZETAneu}
\ee
where we have introduced
\beq
\begin{array}{c}
\ds\int_0^{\infty} dx \, x^{-s-1} \, \left\{
\ln\left[ {2m\over \sqrt{1+x^2}} x^2 I_{m}'(m x) K_{m}'(m x) \right]
-2\sum_{2 \le 2n \le N} { {\cal U}^{{\mbox{\ssz }} {\cal N}}_{2n}(t(x)) \over m^{2n} }
\right\}
\equiv {\cal S}^{{\cal N}}_N(s, m) .
\end{array}
\eeq
and \cam
\begin{equation}
\begin{array}{lll}
\ds{\cal U}^{{\mbox{\ssz }} {\cal N}}_2(t)&=&\ds
-{{3{t^2}}\over {16}} + {{5\,{t^4}}\over 8} -
 {{7\,{t^6}}\over {16}}, \\
\ds{\cal U}^{{\mbox{\ssz }} {\cal N}}_4(t)&=&\ds
    -{{27\,{t^4}}\over {128}} +
     {{109\,{t^6}}\over {32}} -
     {{733\,{t^8}}\over {64}} +
     {{441\,{t^{10}}}\over {32}} -
     {{707\,{t^{12}}}\over {128}}  \\
&\vdots&
\end{array}
\label{U14N}
\end{equation}
In order to obtain the complete Neumann $\zeta$ function we need the
sum
\ba
\sum _{m=1} ^{\infty}  \zeta ^{\cal N}_m (s)
&=& {1 \over 2 } \zeta _R(s)  + {s^2 (2+7s)\over 128 } \zeta _R(s+2)
+{s^2(s+2)(-16 +548 s +2156 s^2 +707 s^3) \over 491520 } \zeta _R(s+4)
\nonumber \\
&+&{s \over \pi} \sin \bigg( {\pi s \over 2} \bigg)  
\sum _{m=1} ^{\infty} m^{-s}{\cal S}^{\cal N}_4(s,m) 
\label{sumZETAneu}
\ea
%
%
%
%
%
%
As we content ourselves with $N=4$, we will only need
${\cal S}_4^{\cal N}(-2,m)$,
After some calculations, we have obtained:
{\small
\ba
&& {\cal S}_4^{\cal N}(-2,1) = -2.3057255 \times 10 ^{-3}, \quad
{\cal S}_4^{\cal N}(-2,2) = -4.755908 \times 10 ^{-5}, \quad
{\cal S}_4^{\cal N}(-2,3) = -4.495047 \times 10 ^{-6}, \nonumber \\
&& {\cal S}_4^{\cal N}(-2,4) = -8.2371882 \times  10 ^{-7}, \quad
{\cal S}_4^{\cal N}(-2,5) = -2.1903952 \times  10 ^{-7}, \quad
{\cal S}_4^{\cal N}(-2,6) = -7.3944610 \times  10 ^{-8}, \nonumber \\
&& {\cal S}_4^{\cal N}(-2,7) = -2.9468381 \times  10 ^{-8}, \quad
{\cal S}_4^{\cal N}(-2,8) = -1.3267927 \times  10 ^{-8}, \quad \cdots
\ea
}
For sufficiently large $m$, ${\cal S}_4^{\cal N} (-2,m)$ can be approximated by
\be
{\cal S}_4^{\cal N} (-2,m) = -{9 \over 2560 m^6}
+{1657 \over 675840 m^8}
-{494891 \over 147603456 m^{10}}
+{10472667 \over 1394032640 m^{12}} ,
\ee
which implies that 
\be
\sum_{m=1}^\infty m^2 {\cal S}_4^{\cal N} (-2,m) =
-0.00256191227
\ee
Again,  \req{theZETAneu} is not valid for $m = 0$, which has to be
obtained separately:
\be
\zeta_0^{\cal N}(s)= {3 s \over 16} +
{s \over \pi} \sin \bigg( {\pi s \over 2} \bigg)
\int _0 ^\infty x^{-1-s} dx \, \bigg\{
\ln \left[ -2 x I_0'(x) K_0'(x) \right]
+{3 \over 8} {1 \over 1+x^2} \bigg\} ,
\ee
which for $s \sim -2$ can be approximated by
\be
\begin{array}{lll}
\zeta_0^{\cal N}(s)&=&\ds {3 s \over 16} 
+{s \over \pi} \sin \bigg( {\pi s \over 2} \bigg)
[ -0.475214928727027
 + {\cal O}(s+2) ] \\
&=&\ds -{3 \over 8}+\left( {3 \over 16} 
-0.475214928727027
\right)(s+2)
+ {\cal O}\left( (s+2)^2 \right) .
\end{array}
\ee
Gathering everything together, we find that the contribution of the
Neumann modes is 
\be
\varepsilon^{\cal N} = -{ 0.01417613719   \over a^2} .
\label{laN}
\ee
which, as expected, is free of divergences.
Here $\varepsilon^{\cal N}$ means the part of $\epsilon$ coming from
the contribution of $\zeta_{\cal M}^{\cal N}$ alone. It can be also viewed
as the Casimir energy linear density of a scalar field subject
Neumann b.c. on the cylindrical surface.

\lsect{Result and discussion}
The total electromagnetic energy per unit of length is given by
the sum of \req{scalarENE} and \req{laN}, which yields
\be
\varepsilon \equiv \varepsilon^{\mbox{\ssz e.m.}} 
= \epsilon^{\cal D}+\epsilon^{\cal N}=
-{0.013561343 \over a^2}
\label{totalDEN}
\ee
\cam
which confirms the result obtained in ref. \cite{MiRa}, where a
high-frequency cutoff procedure was used, as well as our new 
results \req{scalarENE} and \req{laN}.
The result in \req{totalDEN} is negative and indicates
an ensuing attractive force. However, only the Neumann modes give
a contribution of the same sign as the net result. 
The part from the Dirichlet modes
is two orders of magnitude smaller and has opposite sign.
In view of the analogous problem for the circle in $2+1$ dimensions
\cite{LRap},
where the e.m. modes reduce to a set of Neumann modes only,
this dominance is perhaps not so surprising. Nevertheless, in that case
the result was infinite before applying the principal part prescription
(by zeta-function regularization, the type of divergence involved is seen 
to disappear when the number of space dimensions is odd).

Curiously enough, one of the initial motivations for calculating
$\varepsilon^{\mbox{\ssz e.m.}}$ was the conjecture \cite{BaDu} that, being an
intermediate situation between the parallel plates (attractive) and
the sphere (repulsive), the cylinder might turn out to have 
$\varepsilon^{\mbox{\ssz e.m.}}=0$.
What we see is that the two contributions to the e.m. energy
do indeed have opposite signs, although they fail to cancel each other
as the authors of that speculation might have hoped long ago.

In ref.\cite{NP} the scalar field under Neumann b.c. was separately studied
for the spherical case. Using the figure in that paper 
(and the ones in previous works for the other cases with a sphere), we can
compare the Casimir energy per area unit $E_C/A$ for the sphere and for
an infinite cylinder with the same radius ($a$):
\begin{center}
\begin{tabular}{|c|c|c|c|}
\hline
$a^3 \cdot E_C/A$&${\cal D}$-scalar&${\cal N}$-scalar&e.m. \\ \hline
cylinder& $ +0.000098 $ & $ -0.002256 $ & $ -0.002158 $ \\
sphere  & $ +0.000224 $ & $ -0.017808 $ & $ +0.003689 $ \\
\hline
\end{tabular}
\end{center}
Observe that, for the sphere, the e.m. case is {\it not} ${\cal D}+{\cal N}$
because, in the presence of that surface, it decomposes into Dirichlet 
plus some special form of Robin ---not Neumann--- modes.
Actually, the rate between the ${\cal D}$ and ${\cal N}$ contributions
to the cylinder is $-0.043\dots$.
In the spherical case, the ratio of `${\cal D}$ to ${\cal N}$'
has the value of $-0.012\dots$, which has the same sign and order
of magnitude, but is nearly four times smaller. Note that, in the
case of the parallel plates, this ratio is $+1$.
Comparing the e.m. results alone, the sphere has almost twice the absolute value
of the cylinder.
At any rate, since the energy for the cylinder is lower, the Casimir
effect might tend to deform conducting spherical bubbles into
cylindrical tubes. In a `foam-like' universe model, one could even
imagine a bubble becoming very thin in some directions and very long
in another, as was conjectured in ref.\cite{AW}.

Regarding the two parts as separate results for scalar fields under
different b.c., we observe that a change in the nature of the
conditions can produce a drastic alteration, in size and sign, of
the vacuum energy.
Thus, the Casimir effect for a scalar field under
Dirichlet b.c. would tend to expand the cylinder, but, after changing
them into Neumann b.c., the effect would tend to make the cylinder 
contract.
This is, too, the tendency of the e.m. field, as was duly pointed out by the
authors of ref.\cite{MiRa}. In the same paper, they also said:
{\it Still eluding us is the physical intuition to predict
Casimir attraction or repulsion}. Seventeen years later, the truth of this
remark seems to hold on.

\vskip3ex
\lsect{Acknowledgements}
A.R. is grateful to {\it Generalitat de Catalunya ---Comissionat per a
Universitats i Recerca} for financial support to make a stay, in summer 1997, in
Dept of Physics and Astronomy, University of Oklahoma (1996BEAI300119),
where some of the ideas in this paper were conceived.
Thanks are also due to R. Kantowksi and to K.A. Milton himself, 
for the discussions which took place there. S. Leseduarte is kindly thanked
for bibliographic help.

\end{document}